\documentclass{IOS-Book-Article}
\usepackage[utf8]{inputenc}
\usepackage[T1]{fontenc}
\usepackage{times}
\usepackage{tikz}
\usepackage{booktabs}
\usepackage{tabularx}
\usepackage{mathtools,amsthm}
\theoremstyle{definition}
\newtheorem{ex}{Example}

\newcommand\pif{\,{\mid}\,}
\newcommand\apprx{}\let\apprx\tilde
\newcommand\rank{\operatorname{rank}}
\newcommand\similarity{\operatorname{similarity}}
\begin{document}
\begin{frontmatter}

\title{Dermtrainer%
\thanks{Supported by the Austrian Research Promotion Agency (FFG), grant 840281.}}
\runningtitle{Dermtrainer}
\subtitle{A Decision Support System for Dermatological Diseases}
\author[A]{\fnms{Gernot} \snm{Salzer}%
	\thanks{Corresponding Author: Gernot Salzer, Technische Universität
		Wien, Karlsplatz~13/E185-2, 1040 Wien, Austria; E-mail:
		gernot.salzer@tuwien.ac.at.}},
\author[A]{\fnms{Agata} \snm{Ciabattoni}},
\author[A]{\fnms{Christian} \snm{Fermüller}},
\author[B]{\fnms{Martin} \snm{Haiduk}},
\author[C]{\fnms{Harald} \snm{Kittler}},
\author[B]{\fnms{Arno} \snm{Lukas}},
\author[A]{\fnms{Rosa María} \snm{Rodríguez Domínguez}},
\author[C]{\fnms{Antonia} \snm{Wesinger}},
and
\author[C]{\fnms{Elisabeth} \snm{Riedl}}
\runningauthor{G.\ Salzer et al.}
\address[A]{Technische Universität Wien, Vienna, Austria}
\address[B]{Emergentec Biodevelopment GmbH, Vienna, Austria}
\address[C]{Medizinische Universität Wien, Vienna, Austria}

\begin{abstract}
  Dermtrainer is a medical decision support system that assists general practitioners in diagnosing skin diseases and serves as a training platform for dermatologists.
  Its key components are a comprehensive dermatological knowledge base, a clinical algorithm for diagnosing skin diseases, a reasoning component for deducing the most likely differential diagnoses for a patient, and a library of high-quality images.
  This report describes the technical components of the system, in particular the ranking algorithm for retrieving appropriate diseases as diagnoses.
\end{abstract}
\end{frontmatter}
\section{Motivation}

Skin diseases, most prominently skin cancer, pose a significant burden on public health, with incidences continually increasing~\cite{lomas2012}.
Early recognition and knowledge about the epidemiology of the various skin diseases are prerequisites to treat them effectively and to prevent fatal outcomes~\cite{gershenwald2017}.
These necessities face two major obstacles: In many areas of the world dermatology specialists are rare, and non-specialists are responsible for diagnosing skin diseases.
Moreover, within Europe, training programs for dermatologists vary substantially between countries.

Available diagnostic decision support tools for dermatology often perform simple data retrieval without any reasoning component~\cite{tschandl2018cbir}, are cumbersome to use, frequently yield poor results~\cite{delrosario2018}, and lack scientific validation~\cite{ngoo2018}.
The objective of the project \emph{Dermtrainer} (funded by FFG, 2013--2015) was to develop a medical decision support system that serves as a training platform for dermatologists and assists general practitioners in diagnosing skin diseases.

\section{Consortium}

\emph{Dermatologists} from the Medical University of Vienna~-- headed by Elisabeth Riedl who was also overall coordinator~-- contributed a clinical algorithm (a systematic procedure for diagnosing skin diseases) and entered data describing skin diseases into a knowledge databases.
They were also in charge of a clinical validation study comparing the diagnostic accuracy of non-specialized physicians when using Dermtrainer vs.\ a standard encyclopedia (manuscript in preparation at time of submission).

\emph{Software engineers} from the company Emergentec Biodevelopment~-- coordinated by Arno Lukas~-- developed the user interface as well as the infrastructure for storing and maintaining data and images.
Moreover, they provided computer support for the clinical validation study.

\emph{Computer scientists} from Technische Universität Wien~-- coordinated by Gernot Salzer~-- analyzed the meaning of the data, devised, implemented, and evaluated methods for selecting appropriate diagnoses based on patient data, and provided a tool that helped in debugging the disease data.

\section{User interface}

\begin{figure}
  \centering
  \pgfimage[height=4cm]{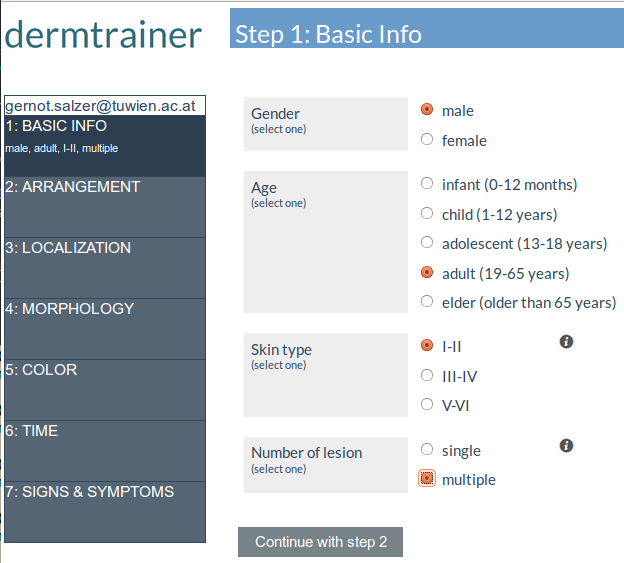}\hspace{-2cm}
  \raisebox{-0.5cm}{\pgfimage[height=4cm]{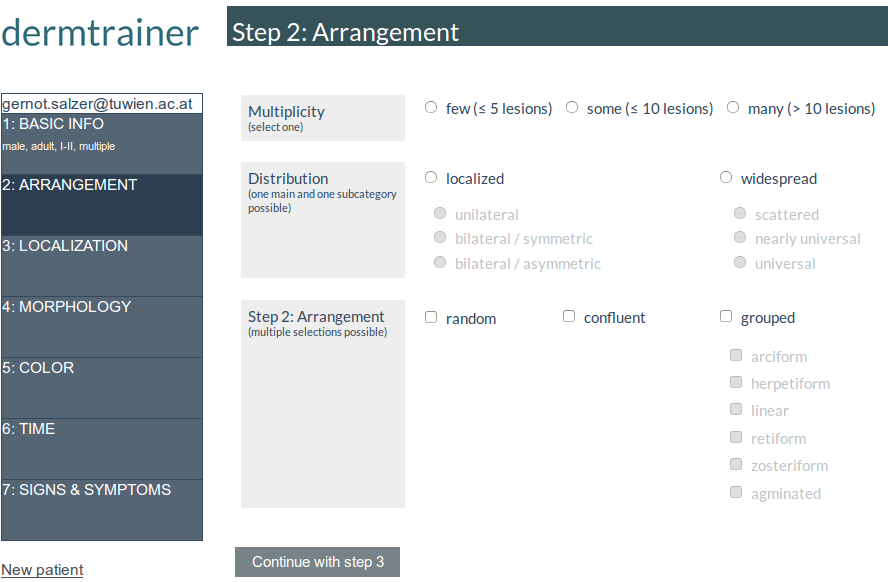}}\hspace{-2cm}
  \raisebox{-1cm}{\pgfimage[height=4cm]{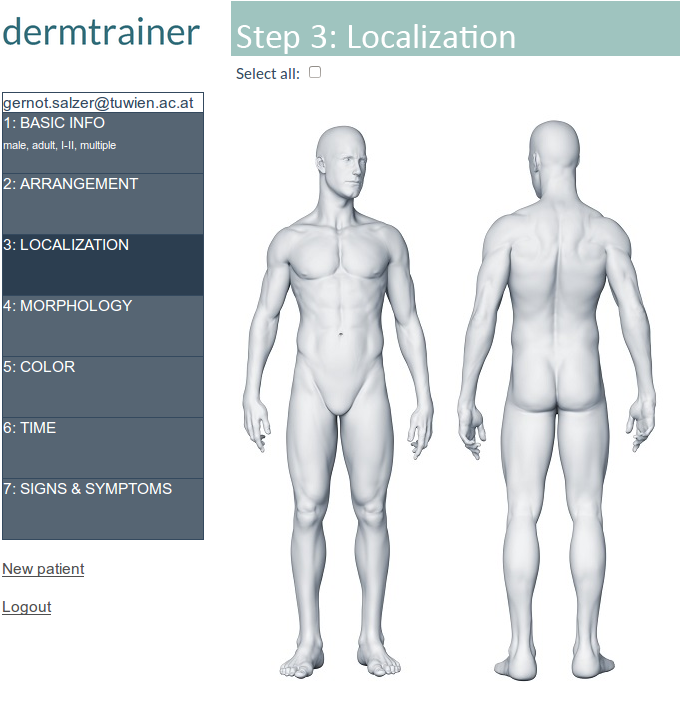}}
  \pgfimage[height=4cm]{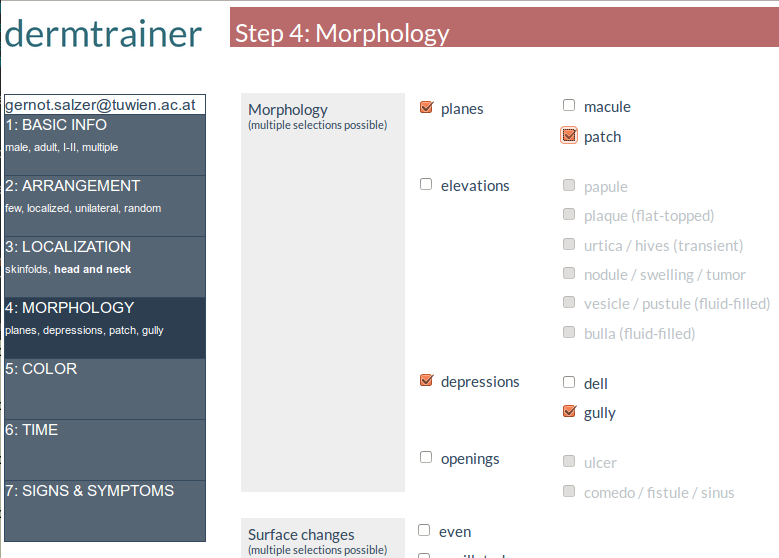}\hspace{-3.5cm}
  \raisebox{-1cm}{\pgfimage[height=4cm]{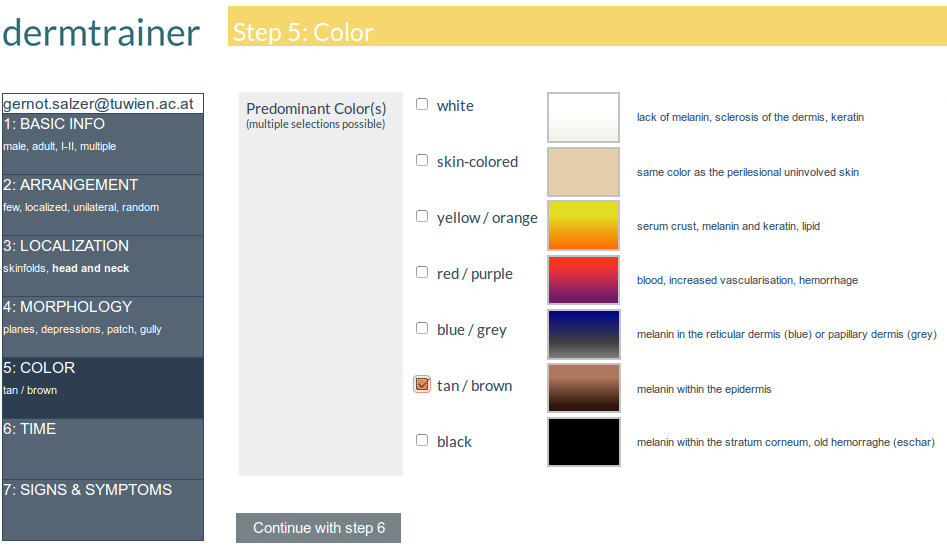}}\hspace{-2.5cm}
  \raisebox{-0.5cm}{\pgfimage[height=4cm]{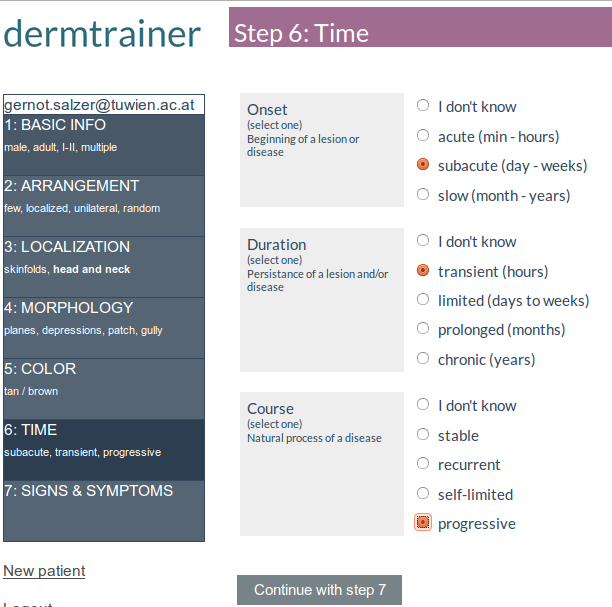}}\hspace{-3cm}
  \raisebox{-1cm}{\pgfimage[height=4cm]{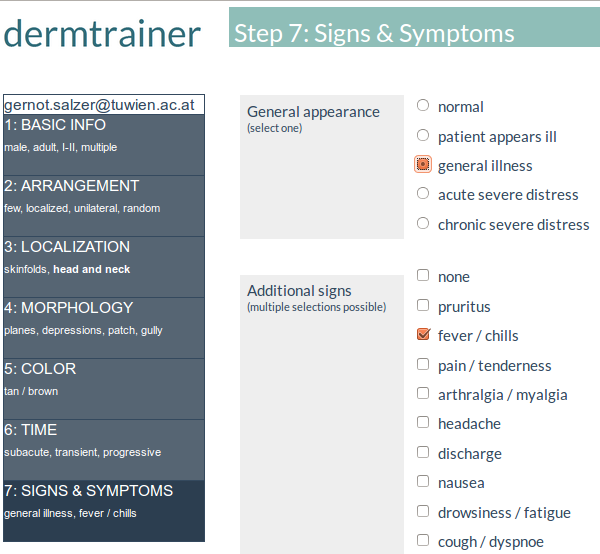}}
  \caption{The patient data is entered in seven steps, starting with
    basic information like age and sex, continuing with arrangement,
    morphology, location, and color of lesions, and concluding with
    timing information and additional signs.}
  \label{fig:steps}
\end{figure}

Physicians enter data of a patient case via a web interface in seven steps, providing information on the arrangement, localization, morphology, and color of the observed lesions as well as timing information and additional signs (Figure~\ref{fig:steps}).
It is not necessary to fill the forms completely, but more input data potentially leads to more accurate rankings.

\section{Knowledge Database}

The knowledge database (at the time of conclusion of the Austrian Research Promotion Agency (FFG) project) contains the data of 620 diseases. Each is represented by 150 fields, of which 133 correspond to symptoms that are used for ranking (Table~\ref{tab:values}) and 17 provide supplementary information.
\begin{table}
  \centering
  \begin{tabular}{lrl}
    \toprule
    type & fields & examples\\
    \midrule
    number \& arrangement & 22 & single/multiple, localized/widespread\\
    location & 17 & neck, arm, abdomen\\
    morphology &37 & papule, ulcer, arciform\\
    color & 16 & single/multiple, white, blue\\
    timing & 11 & acute, chronic, recurrent\\
    demographics & 8 & age groups, skin type\\
    signs \& symptoms & 19 & general appearance, additional signs\\
    epidemiology & 3 & frequency, gender ratio\\
    \bottomrule
  \end{tabular}
  \caption{Overview of the fields relevant for ranking a disease}
  \label{tab:values}
\end{table}

\subsection{Likelihood}

Most values in the database approximate the \emph{conditional probability} $P(s\pif d)$ that the symptom~$s$ is observed in the case that the patient has the disease~$d$.
In a perfect world we would know this probability for each symptom and each disease.
In reality, virtually none of the more than 80\,000 values can be determined with absolute certainty.
They are replaced by one of three judgements, estimated by the dermatologists: ``yes, this symptom occurs with the disease'', ``no, it does not'', and ``this symptom is unlikely to occur, but it may'' (Table~\ref{tab:likelihood}).
\begin{table}
  \centering
  \begin{tabularx}{\textwidth}{@{}lXc@{}}
    \toprule
    judgement & meaning & likelihood $L(s,d)$\\
    \midrule
    yes        & The symptom occurs with the disease. & 1.0\phantom{00}\\
    unlikely   & In rare cases the symptom can be observed with the disease. & 0.02\phantom{0} \\
    no         & The symptom never occurs with the disease, but the disease should remain in the ranking. & 0.001\\
    no         & The symptom never occurs with the disease, the disease can be excluded from ranking. & 0.0\phantom{00} \\
    \bottomrule
  \end{tabularx}
  \caption{%
    Judgements regarding whether a symptom may occur with a disease.
    The likelihood is used later on for ranking.}
  \label{tab:likelihood}
\end{table}
For the purpose of ranking, we represent the judgements by numerical values that we call \emph{likelihood} (to avoid the term probability); $L(s,d)$ denotes the likelihood that the symptom~$s$ occurs with disease~$d$.
This fuzzy knowledge is one of the main challenges for automated diagnosis.

\subsection{Simplified Likelihood}

For demographics (predominant age group and skin type) and location (body sites where lesions occur) the judgement ``unlikely'' is not used.
The database only contains information about whether or not the disease occurs among persons having a specific skin type or age, or whether the lesions are confined to specific sites of the body.
There is no distinction between ``yes'' and ``unlikely''.

\subsection{Epidemiology}

There are a few elements with higher precision, most notably the overall frequency of the disease and the sex ratio, extracted from literature and expert knowledge.
The frequency is specified on a 6-level scale from ``exceptional'' to ``very common'' (Table~\ref{tab:frequency}).
\begin{table}
  \centering
  \begin{tabular}{ll}
    \toprule
    frequency   & $F(d)$ \\
    \midrule
    exceptional & $10^{-7}$ \\
    rare        & $10^{-6}$ \\
    uncommon    & $10^{-5}$ \\
    less common & $10^{-4}$ \\
    common      & $10^{-3}$ \\
    very common & $10^{-2}$ \\
    \bottomrule
  \end{tabular}
  \caption{Frequency of diseases}
  \label{tab:frequency}
\end{table}
If a disease is known to affect one sex more frequently than the other, it is possible to specify a $\textrm{male}\,{:}\,\textrm{female}$ ratio.
E.g., the ratio $2\,{:}\,1$ indicates that the disease occurs twice as often among men, while $0\,{:}\,1$ indicates that the disease occurs exclusively among women.

\section{Ranking}

Diseases are ranked in three phases.

\subsection*{Phase 1: Exclusion of Diseases}

The number of diseases is reduced to typically less than a hundred by eliminating diseases where certain symptoms selected in the user interface correspond to a ``no'' value in the disease record:
If a particular symptom never occurs with a disease, then the disease can be excluded as a potential diagnosis.

Unfortunately, this approach only works for some symptoms that cannot be mistaken.
In preliminary testing, we found that judgements of non-specialists may differ considerably from the opinion of dermatologists who defined the disease specifications, leading to spurious exclusion of diseases by mistaking single symptoms.

We therefore distinguish between exclusive and non-exclusive symptoms.
\emph{Exclusive} symptoms are those that can be classified beyond doubt.
For such symptoms the judgement ``no'' will exclude the disease from the list of diagnoses.
An example for an exclusive symptom is the number of lesions: If the patient shows just a single lesion, diseases that always go with multiple lesions are excluded from further consideration.
For \emph{non-exclusive} symptoms ``no'' is not a strict ``no'' but rather a ``even more unlikely than unlikely''.
The disease is penalized, but stays in the ranking and may still become a top diagnosis if other symptoms make up for the penalty.

In most cases, a symptom is exclusive for all diseases or for none.
Locations are an exception, as their exclusiveness can be specified independently per disease.
It is possible to specify for a disease that the judgement ``no'' for a location like the head is to be interpreted strictly:
If the patient shows a lesion there, the disease is excluded as diagnosis.
Other diseases may handle the location more liberally and just downgrade a disease if the observed lesion is in the wrong place.

\subsection*{Phase 2: Scoring}

In the second phase we compute scores that reflect how well each of the non-excluded diseases matches the provided symptoms.
After evaluating several approaches, including term frequency-inverse document frequency (tf-idf)~\cite{manning2008introduction} many-valued logics, we finally settled on a method that is based on a probabilistic interpretation of the data.
It yields two numbers per disease, one called \emph{similarity} that ignores epidemiology, and one called \emph{rank} that combines similarity with the overall frequency of the disease.
The details of the method are presented in sections \ref{sec:bayes} and~\ref{sec:approx}.

\subsection*{Phase 3: Selection}

\begin{figure}
  \centering
  \pgfimage[height=6cm]{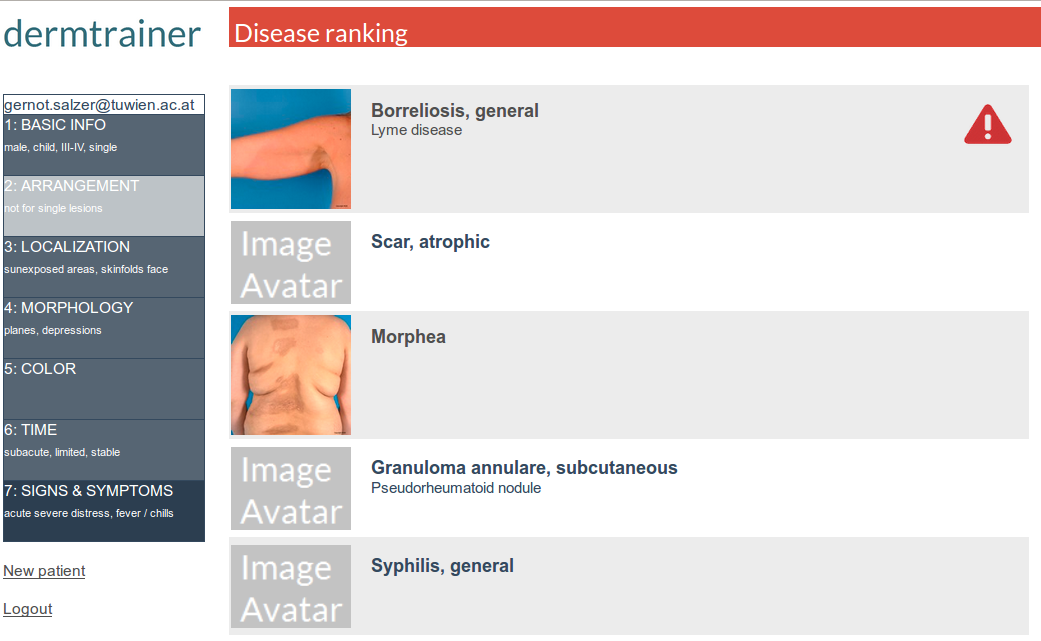}
  \caption{Sample view of a disease ranking after entering symptom attributes.
    Warning signs mark diseases prespecified as severe.}
  \label{fig:ranking}
\end{figure}

In the third phase we select the diseases with the highest scores regarding similarity and rank (above a fixed threshold) and display them as the possible diagnoses.
Even though the diagnoses are displayed in the order of their internal ranking, we avoid to show ranks and scores, as these numbers would convey a wrong sense of precision, and might be mistaken as actual probabilities (Figure~\ref{fig:ranking}).

\section{A Ranking Method Based on Bayes' Theorem}\label{sec:bayes}

We use a method sometimes referred to as naïve Bayes classification, which relies on Bayes' theorem with strong independence assumptions~\cite{HandYu2001}.

Given a patient with particular symptoms, the aim is to assign a rank to each disease that indicates how well the symptoms are explained by the disease.
One approach is to use $P(d\pif s_1,\dots,s_n)$, the conditional probability of disease~$d$ given the symptoms $s_1, \dots,s_n$, as the rank of~$d$.
According to Bayes' theorem it can be computed as
\begin{align}
P(d\pif s_1,\dots,s_n)
&= \frac{P(s_1,\dots,s_n\pif d)\cdot P(d)}{P(s_1,\dots,s_n)}
\label{eq:bayes}
\end{align}
where
\begin{flushleft}
  \begin{tabularx}{\textwidth}{@{}lX@{}}
    $P(d\pif s_1,\dots,s_n)$ & probability of the patient having disease~$d$ if showing the symptoms $s_1,\dots,s_n$\\
    $P(s_1,\dots,s_n\pif d)$ & probability of observing all of the symptoms $s_1,\dots,s_n$ simultaneously if the patient has disease~$d$\\
    $P(d)$ & probability of disease~$d$\\
    $P(s_1,\dots,s_n)$ & probability of observing all of the symptoms $s_1,\dots,s_n$ simultaneously (caused by whatever disease)
  \end{tabularx}
\end{flushleft}
Note that we are not interested in the probability $P(d\pif s_1,\dots,s_n)$ itself but rather in the relative order of diseases with respect to this value.
This order is maintained when multiplying the probability with a constant.  $P(s_1,\dots,s_n)$ is such a constant since it is independent of any particular disease.
Therefore we define the rank $\rank(d\pif s_1,\dots,s_n)$ of a disease~$d$ for symptoms $s_1,\dots,s_n$ as
\begin{align}
\rank(d\pif s_1,\dots,s_n)
&= P(d\pif s_1,\dots,s_n)\cdot P(s_1,\dots,s_n)\\
&= P(s_1,\dots,s_n\pif d)\cdot P(d)
\label{eq:rank1}
\end{align}
We assume the symptoms to be independent from each other.
This allows us to replace $P(s_1,\dots,s_n\pif d)$ by $P(s_1\pif d)\cdots P(s_n\pif d)$.
Strictly speaking most symptoms are not independent, but the correlation is weak.
The simplification is also justified \emph{ex post} by the good results of the final evaluation of the algorithm.
Thus equation~\eqref{eq:rank1} can be rewritten as
\begin{align}
\rank(d\pif s_1,\dots,s_n)& = P(s_1\pif d)\cdot\cdots\cdot P(s_n\pif d)\cdot P(d)
\label{eq:rank2}
\end{align}
where
\begin{flushleft}
  \begin{tabularx}{\textwidth}{@{}lX@{}}
    $\rank(d\pif s_1,\dots,s_n)$ & rank of disease~$d$ given the symptoms
    $s_1,\dots,s_n$\\
    $P(s_i\pif d)$ & conditional probability of observing symptom
    $s_i$ given disease~$d$\\
    $P(d)$ & unconditional probability for disease~$d$
  \end{tabularx}
\end{flushleft}
Determining precise values for $P(s_i\pif d)$ and $P(d)$ is virtually impossible, as it would require extensive medical studies for all combinations of symptoms and diseases.
Therefore we use approximations $\apprx P(s_i\pif d)$ and $\apprx P(d)$ derived from the qualitative information in the database.
\begin{align}
\rank(d\pif s_1,\dots,s_n)& = \apprx P(s_1\pif d)\cdot\cdots\cdot \apprx P(s_n\pif d)\cdot \apprx P(d)
\label{eq:rank3}
\end{align}
Some symptoms may turn out to be more relevant for diagnosing than others.
This can be modeled by attaching weights to the symptoms; in a multiplicative formula this means to add exponents.
\begin{align}
\rank(d\pif s_1,\dots,s_n)& = \apprx P(s_1\pif d)^{w_1}\cdot\cdots\cdot \apprx P(s_n\pif d)^{w_n}\cdot \apprx P(d)
\label{eq:rank4}
\end{align}
The first part of expression~\eqref{eq:rank4} quantifies how well the disease fits the symptoms, regardless of its frequency.
We call this part \emph{similarity}. This leads us to the final formula that serves as the basis of the ranking algorithm.
\begin{align}
  \rank(d\pif s_1,\dots,s_n)& = \similarity(s_1,\dots,s_n\pif d)\cdot \apprx P(d) \label{eq:rank5}\\
  \similarity(s_1,\dots,s_n\pif d) &= \apprx P(s_1\pif d)^{w_1}\cdot\cdots\cdot \apprx P(s_n\pif d)^{w_n}
\label{eq:rank6}
\end{align}
where
\begin{flushleft}
  \begin{tabularx}{\textwidth}{@{}lX@{}}
    $\rank(d\pif s_1,\dots,s_n)$ & weighted rank of disease~$d$ given the symptoms $s_1,\dots,s_n$\\
    $\similarity(s_1,\dots,s_n\pif d)$ & similarity between symptoms $s_1,\dots,s_n$ and disease~$d$\\
    $\apprx P(s_i\pif d)$ & approximate probability of observing symptom $s_i$ given disease~$d$\\
    $\apprx P(d)$ & approximate probability of disease~$d$\\
    $w_1,\dots,w_n$ & weights of the symptoms $s_1,\dots,s_n$; choose $w_1=\cdots=w_n=1$ for no weighting
  \end{tabularx}
\end{flushleft}

\section{From Judgements to Approximate Probabilities}\label{sec:approx}

In this section we discuss how to obtain the approximations $\apprx P(s_i\pif d)$ and $\apprx P(d)$ in equations \eqref{eq:rank5} and \eqref{eq:rank6} from the disease database.

\subsection{Categories of Symptoms}

A \emph{category} is a group of symptoms that are considered simultaneously.
The symptoms of a category may be mutually exclusive, meaning that the user can select at most one.
If several ones can be selected, their average is computed in order to give equal weight to the categories, independent of the number of symptoms they are comprised of and the number of symptoms that have been selected.
The above mentioned constraints for the selection of symptoms as well as dependencies between categories are hard-wired into the user interface and are automatically enforced.

\begin{ex}
  The category \emph{predominant age group} contains the five `symptoms' \emph{infant}, \emph{child}, \emph{adolescent}, \emph{adult}, and \emph{elder}, of which only one can be selected.
  Likewise, the symptoms \emph{localized} and \emph{widespread} in the category \emph{distribution} are mutually exclusive.
\end{ex}
\begin{ex}
  In category \emph{form}, any number of the symptoms \emph{domeShaped}, \emph{flatTopped}, and \emph{umbilicated} may be selected simultaneously.  
\end{ex}

\subsection{Computing the Approximate Probability of a Category}

Let $s_1,\dots,s_n$ be all symptoms of a category, and let $t_1,\dots,t_k$ be the symptoms actually observed.
The approximate probability for these observations under the assumption that the patient has disease~$d$, denoted by $\apprx P(t_1,\dots,t_k\pif d)$, is obtained by the following formula.
\begin{align}
  \apprx P(t_1,\dots,t_k\pif d) &= \frac{\sqrt[k]{L(t_1,d)\cdot\cdots\cdot L(t_k,d)}}{L(s_1,d)+\cdots+L(s_n,d)}
\end{align}
Multiplying these approximate probabilities for all categories of a disease yields its similarity with respect to the observed symptoms (compare equation~\eqref{eq:rank6} above).

\begin{ex}
  Consider the symptoms $s_1=\text{\emph{domeShaped}}$, $s_2=\text{\emph{flatTopped}}$, and $s_3=\text{\emph{umbilicated}}$ of category \emph{form}.
  Suppose the patient shows the symptoms $t_1=\mathit{domeShaped}$ and $t_2=\mathit{umbilicated}$.
  For the disease $d=\text{\emph{Atypical fibroxanthoma}}$ the expert judgement for the symptoms \emph{domeShaped} and \emph{flatTopped} is `yes', whereas \emph{umbilicated} is considered `unlikely'.
  The corresponding likelihoods are $L(s_1,d)=L(t_1,d)=L(s_2,d)=1.0$ and $L(s_3,d)=L(t_2,d)=0.02$.
  Hence we obtain:
  \begin{align}
    \MoveEqLeft\apprx P(\text{\emph{domeShaped}},\text{\emph{umbilicated}}\pif \text{\emph{Atypical fibroxanthoma}})&\\
    &= \frac{\sqrt[2]{L(t_1,d)\cdot L(t_2,d)}}{L(s_1,d)+L(s_2,d)+L(s_3,d)}\\
    &= \frac{\sqrt[2]{1.0\cdot 0.02}}{1.0+1.0+0.02}\\
    &= 0.07
  \end{align}
\end{ex}

\subsection{Epidemiology}

To compute the rank of a disease, we have to multiply its similarity with its probability (equation~\eqref{eq:rank5}).
The probability $P(d)$ is approximated by the frequency of the disease in Table~\ref{tab:frequency}.
\begin{align}
  \apprx P(d)&= F(d)
\end{align}

\section{Evaluation}

Evaluation was done in three stages.
First, dermatologists participating in the project tested the tool with retrospective cases from their practice for plausibility of results.
Unreasonable rankings were analyzed, leading to further modifications in the database and improvements of the ranking algorithms.

For the subsequent evaluations, virtual patient cases were created consisting of the most common and important skin diseases.
The second stage was performed on dermatology residents of a single center to test the feasibility of the system, and the third stage on non-dermatology residents (publications pending).

One limitation of the evaluations so far has been the necessity to rely on virtual patients for ethical reasons.
Symptoms like the three-dimensional form~-- plane, elevation, depression, opening~-- or the consistency~-- soft, firm, indurated~-- are hard to judge from an image.
A dermatologist can make up for it by matching the images with real cases from the past, but non-experts, one of the target groups, will probably often guess wrongly.
We expect that Dermtrainer will perform better with real patients, when such symptoms can be correctly determined.

\bibliographystyle{plain}
\bibliography{dermtrainer}

\end{document}